\documentclass[preprint, iop]{emulateapj}

\usepackage{color}
\usepackage{enumitem}
\usepackage[fleqn]{amsmath}
\usepackage{bm}
\usepackage[breaklinks,colorlinks,urlcolor=blue,citecolor=blue,linkcolor=blue]{hyperref}

\shorttitle{Non-Transiting Hot Jupiters}
\shortauthors{Millholland et al.}

\begin{document}
\setlength{\mathindent}{0.0pt}

\title{On the Detection of Non-Transiting Hot Jupiters in Multiple-Planet Systems}
\author{Sarah Millholland\altaffilmark{1}, Songhu Wang\altaffilmark{1,2}, Gregory Laughlin\altaffilmark{1}}
\affil{$^{1}$Department of Astronomy and Astrophysics, University of California at Santa Cruz, Santa Cruz, CA 95064}
\affil{$^{2}$School of Astronomy and Space Science and Key Laboratory of Modern Astronomy and Astrophysics in Ministry of Education, Nanjing University, Nanjing 210093, China}
\email{smillhol@ucsc.edu}

\begin{abstract}
We outline a photometric method for detecting the presence of a non-transiting short-period giant planet in a planetary system harboring one or more longer period transiting planets. Within a prospective system of the type that we consider, a hot Jupiter on an interior orbit inclined to the line-of-sight signals its presence through approximately sinusoidal full-phase photometric variations in the stellar light curve, correlated with astrometrically induced transit timing variations for exterior transiting planets. Systems containing a hot Jupiter along with a low-mass outer planet or planets on inclined orbits are a predicted hallmark of in situ accretion for hot Jupiters, and their presence can thus be used to test planetary formation theories. We outline the prospects for detecting non-transiting hot Jupiters using photometric data from typical \textit{Kepler} objects of interest (KOIs). As a demonstration of the technique, we perform a brief assessment of \textit{Kepler} candidates and identify a potential non-transiting hot Jupiter in the KOI-1822 system.  Candidate non-transiting hot Jupiters can be readily confirmed with a small number of Doppler velocity observations, even for stars with $V\gtrsim14$.
\end{abstract}

\section{INTRODUCTION}

Hot Jupiters are both the most readily detectable and the best characterized population of extrasolar planets, yet the dominant mechanism of their formation and evolution remains mysterious. 

Within the most commonly accepted theoretical paradigm, hot Jupiters are thought to form at large radial distances before moving inward \cite[see e.g.][]{2003ApJ...589..605W, 2012ApJ...751..119B, 2012ARAA..50..211K}. Several groups have recently proposed theories \citep{2015arXiv151109157B, 2016ApJ...817L..17B} that contrast with the established ideas of disk migration and suggest that many hot Jupiters form in situ via gas accretion onto 10 - 20 ${M_{\rm {\oplus}}}$ cores. \cite{2015arXiv151109157B} suggest that, under an in situ formation scenario, hot Jupiters should frequently be accompanied by low-mass companions with periods P $\lesssim$ 100 days, sometimes with substantial mutual inclinations. Therefore, the presence or absence of close-in companions to hot Jupiters, with or without mutual inclination, provide a potential zeroth-order test for in situ formation.

Hot Jupiters are thought to largely lack close coplanar planetary companions. This conclusion stems from a paucity of detections of transiting companions to hot Jupiters in \textit{Kepler} data \citep{2011ApJ...732L..24L, 2012PNAS..109.7982S} and a lack of transit timing variations (TTVs) for these objects \citep[e.g.][]{2009ApJ...700.1078G, 2012PNAS..109.7982S}. A recent, K2-facilitated discovery \citep{2015ApJ...812L..18B}, however, of two close companions to the previously discovered hot Jupiter, WASP 47-b, affirms that hot Jupiters can indeed have close planetary companions. Moreover, in a recent search, \cite{2016arXiv160105095H} probed all \textit{Kepler} confirmed and candidate transiting hot Jupiters (P $<$ 10 days) and warm Jupiters (10 $<$ P $<$ 200 days) for transiting companions. While the hot Jupiters have no detectable inner or outer companions with periods P $<$ 50 days and radii R $>$ 2 $ {R_{\rm {\oplus}}}$, about half of the warm Jupiters are closely accompanied by small planets. The authors point to this as evidence supporting an in situ formation scenario for warm Jupiters. 

The transit and TTV search for HJ companions by \cite{2012PNAS..109.7982S} and the transit search by \cite{2016arXiv160105095H} place strong constraints on close-in, coplanar companions, but they are less sensitive to larger period, possibly inclined planets. Furthermore, current radial velocity residuals in systems containing a short-period giant planet are generally at or above the precision required for super-Earth detection. Many low-mass companions to RV-observed hot Jupiters may therefore be lost in the noise.\footnote{Adopting a simple mass-radius relationship, $M_{\rm p}/M_{\oplus}=(R_{\rm p}/R_{\oplus})^2$, one finds that the median radial velocity half-amplitude for \textit{Kepler} candidate planets is $K=1.03\,{\rm m\,s^{-1}}$, whereas the median RMS Doppler residual for currently known hot Jupiters (as tabulated at \url{www.exoplanets.org}) is $\sigma=8.9\,{\rm m\,s^{-1}}$}

Here, we outline a novel technique for detecting \textit{non-transiting} hot Jupiters in systems containing known transiting planets. Our method synergistically combines two well-known detection and characterization strategies: optical phase curve analysis and transit timing variations (TTVs). Specifically, we aim to simultaneously detect measurements of an optical reflection phase curve due to a non-transiting hot Jupiter in conjunction with reflex motion induced (astrometric) TTVs of an outer, transiting, low-mass planet due to the inclined hot Jupiter. The two measurements must be mutually consistent.

In this Letter we detail this ``phase+astrometric TTV" method as a technique to search for non-transiting hot Jupiters in systems containing confirmed or candidate planets. In Section 2, we describe the nature of astrometrically induced TTVs due to a non-transiting hot Jupiter. In Section 3, we review the usage of optical phase curves for the detection and characterization of giant planets. In Section 4, we analyze the detectability of prospective systems. Finally, in Section 5 and Section 6, we describe a brief search of the \textit{Kepler} light curves and present a possible candidate non-transiting hot Jupiter in the KOI-1822 system.

\section{TTVs OF AN OUTER PLANET DUE TO AN INTERIOR, NON-TRANSITING HOT JUPITER}

If an interior, non-transiting hot Jupiter is present in a system hosting a low-mass transiting planet, the HJ will produce small yet detectable non-uniformities in the transiting object's inter-transit times \citep{2005MNRAS.359..567A, 2005Sci...307.1288H}. TTVs of this type are astrometrically induced via the HJ's gravitational effect on the host star and are therefore exceptionally simple to evaluate. 
	
\begin{figure}
\epsscale{1.15}
\plotone{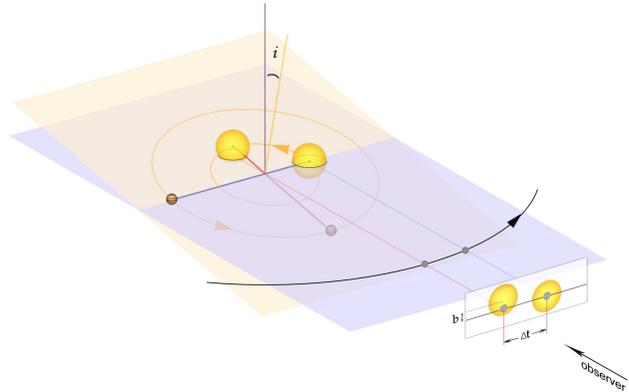}
\caption{A diagram depicting the physical origin of astrometric TTVs. An orbiting planet, shown in gray and orbiting in the light purple plane, transits at different times due to the reflex motion of the star orbiting a non-transiting HJ in the yellow orbital plane.}
\label{AstrometricTTV}
\end{figure}

Consider, for example, a planetary system containing a HJ and an outer, transiting, super-Earth (Figure ~\ref{AstrometricTTV}). The outer planet's contribution to the system's total mass is negligible, so all three bodies may be considered to orbit the star/HJ barycenter. As the HJ orbits, its gravitational influence causes the star to wobble. The outer planet thus orbits a ``moving target'', producing astrometric TTVs fully analogous to those in a circumbinary planetary system \citep{2013MNRAS.434.3047A}.

For the case of coplanar orbits, \cite{2005MNRAS.359..567A} derived expressions for the astrometric TTVs of an outer planet due to an inner perturbing planet with a much smaller periodays. Here we consider the case of a mutual orbital inclination between the planetary orbits. For simplicity we assume an edge-on orbit for the outer planet and a circular, prograde orbit for the inclined HJ.

The TTVs are uniquely specified using the period of the HJ, $P_J$; the HJ's orbital inclination and longitude of the ascending node, $I_J$ and $\Omega_J$; the period of the planet, $P_p$; the eccentricity and argument of periastron of the planet's orbit, $e_p$ and $\omega_p$; the masses, $M_J$ and $M_\star$; the time of the planet's pericenter passage, $t_{\omega_p}$; and the time at which the HJ passes its ascending node, $t_{\Omega_J}$.

The outer planet's mid-transit times occur when the projected distance between the centers of the star and planet is zero. If $e_p\ll1$, such that the guiding center approximation may be used, this amounts to, assuming $M_\star \gg M_J$, 
%
%
\small
\begin{displaymath}
\begin{split}
{n_p}^{\frac{2}{3}}
 \frac{M_J}{M_{\rm{\star}}}
\left(\sin{\Omega_J}\cos\theta_\star(t)
+ \cos{\Omega_J}\sin{I_J}\sin\theta_\star(t)\right)  \\ 
= {n_J}^{\frac{2}{3}}
(1-e_p\cos{M})\sin{(M + 2 e_p\sin{M} + \omega_p)} , 
 \end{split}
\end{displaymath} \normalsize
where $n=2\pi/P$, $\theta_\star(t) = n_J(t - t_{\Omega_J}) + \pi $, and $M = n_p(t-t_{\omega_p})$.


The maximum amplitude of the variations in the inter-transit times is given by (where here it is not necessary that $e_p \ll 1$)
\small
\begin{displaymath}
\Delta T_{\rm TTV}=\frac{1}{\pi}P_p^{\frac{1}{3}}P_J^{\frac{2}{3}}\frac{M_{\rm J}}{M_{\star}}
\frac{[(\sin^2{\Omega_J} + \sin^2{I_J}\cos^2{\Omega_J})(1-e_p^2)]^{\frac{1}{2}}}{1+ e_p \cos{\omega_p}}.
\end{displaymath}
\normalsize

In addition to the timing variations, a non-transiting HJ also induces transit \textit{duration} variations (TDVs) for the outer planet. Under the same assumptions as before, the transit duration is given by\\
\small
\begin{displaymath}
\tau(t) = \left[
\frac{4{R_{\star}}^2}{(v_p - \dot{Y}(t))^2 + \dot{Z}(t)^2} - 
\left(\frac{8 G M_J^3}{n_J^2 M_\star^2}\right)^{\frac{2}{3}}\frac{\cos^2I_J\sin^2{\theta_\star(t)}}{(v_p - \dot{Y}(t))^2}
\right]^\frac{1}{2}
\end{displaymath}
\normalsize 
where
\small
\begin{align*}
\dot{Y}(t) &= \left(\frac{G {M_J}^3 n_J}{{M_{\star}}^2}\right)^\frac{1}{3}(\cos{\Omega_J}\sin{I_J}\cos{\theta_\star(t)}-\sin{\Omega_J}\sin{\theta_\star(t)}), \\
\dot{Z}(t) &= \left(\frac{G  {M_J}^3 n_J}{{M_{\star}}^2}\right)^\frac{1}{3}\cos{I_J}\cos{\theta_\star(t)}, 
\end{align*}
\begin{displaymath}
\text{and  } v_p = (G M_\star n_p)^\frac{1}{3} \frac{1+ e_p \cos{\omega_p}}{(1-{e_p}^2)^\frac{1}{2}}.
\end{displaymath}
\normalsize

The maximum deviation in the transit duration, $\Delta T_{\rm TDV}$, is influenced by both variations in the impact parameter of the transit chord (see Figure ~\ref{AstrometricTTV}) and the relative velocity between the star and planet, and is non-analytic in the general case. When $I_J$ is near $90^\circ$, 
\small
\begin{displaymath}
\Delta T_{\rm TDV}\approx\frac{4 R_{\star}}{(2\pi GM_{\star})^{\frac{1}{3}}}\frac{M_{\rm J}}{M_{\star}}\left(\frac{{P_p}^{2}}{P_{\rm J}}\right)^{\frac{1}{3}}.
\end{displaymath}
\normalsize
For the fiducial case of $M_{\rm J} = M_{\rm Jup}$, $P_{\rm J} = 3$ days, $P_p = 50$ days, $M_{\rm{\star}} = M_{\rm{\sun}}$, $R_{\rm{\star}} = R_{\rm{\sun}}$, and $I_J$ near $90^\circ$, this results in peak-to-peak TTV and TDV amplitudes of 3.4 min and 2.0 min, respectively. These signals are both detectable given a light curve with good enough photometric precision. In the absence of an independent estimate of the HJ's orbital period, however, the prospects for detecting non-transiting HJs using \textit{Kepler}-quality TTVs and TDVs appear bleak. Fortunately, the detection of an optical reflected light phase curve can combine with the TTVs to yield a highly constrained problem. 


\section{OPTICAL PHASE CURVES}


Out-of-transit, optical phase-folded light curves can effectively characterize giant (stellar or sub-stellar) transiting companions \cite[e.g.][]{2011AJ....142..195S, 2013ApJ...772...51E, 2015ApJ...804..150E, 2015AJ....150..112S}. The phase curve, composed of photometry across the out-of-transit orbit, results from the superposition of several independent effects: reflected light and thermal emission, Doppler boosting (beaming) from the reflex motion of the star, and ellipsoidal variations due to tidal forces exerted on the star by the companion \citep{2011AJ....142..195S}. The BEER model \citep{2011MNRAS.415.3921F} is often used to simultaneously analyze these three components. 
For the typical HJ (P $\sim$ 3 days, M $\sim$ 1 ${M_{\rm {Jup}}}$), reflection is the strongest component by up to an order of magnitude. The phase curve in these circumstances is, to first order, sinusoidal. 

While several groups have performed a comprehensive search for phase curve variations in \textit{transiting} HJs \citep[e.g.][]{2012AJ....143...39C, 2013ApJ...772...51E, 2015ApJ...804..150E, 2015PASP..127.1113A}, none have undertaken a thorough search for phase curve detections of \textit{non-transiting} HJs. Authors have frequently discussed the prospects of discovering non-transiting massive planets via their phase curves, for example by performing Bayesian model selection on all phase curve components \citep{2014ApJ...795..112P}. Moreover, phase curves have been used to discover small, non-eclipsing binary stars (0.07 - 0.4 ${M_{\rm {\sun}}}$), in which cases all phase curve components have significant amplitudes \citep{2012ApJ...746..185F}. For the HJs with reflection-dominated phase curves, however, the signals are much more prone to false positives. Fortunately, the simultaneous detection of a reflection-dominated phase curve with astrometric TTVs  breaks this degeneracy.

\section{DETECTABILITY: ANALYSIS OF A FIDUCIAL SYSTEM}

We now consider the detectability of a non-transiting, typical HJ in a fiducial system containing an outer, transiting, low-mass planet. We show that the unseen giant planet is readily detectable given \textit{Kepler}-quality photometry.

Consider a hypothetical planetary system: a 1 $M_{\rm{\sun}}$, 1 $R_{\rm{\sun}}$ host star; a 1 $M_{\rm{Jup}}$, 1.3 $R_{\rm{Jup}}$ HJ on a 3.5 day orbit with geometric albedo $A_{\rm{g}} = 0.2$; and a 3 $R_{\rm{\earth}}$ planet on a circular 80 day orbit. Assume the star is V $\sim$ 13 mag, and assume the HJ's orbit is slightly inclined to the line of sight ($I_J = 83^{\circ}$), yielding a 36 ppm phase curve amplitude.

\begin{figure}
\epsscale{1.2}
\plotone{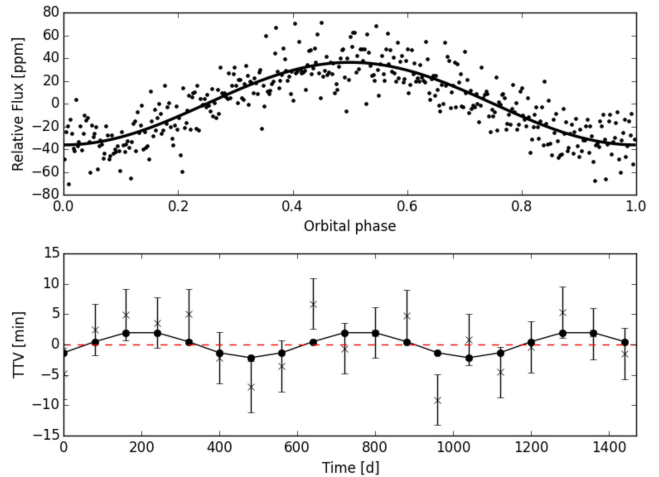}
\caption{The phase curve and TTVs for a fiducial system. The phase curve is folded with period P = 3.5 days and binned such that 400 points span the orbit. The black line indicates a sinusoidal fit. In the TTV plot, the ``x" symbols represent the synthetic data consistent with typical observational uncertainty, and the black dots are the underlying model. }
\label{TestCase10_PhaseTTV}
\end{figure}

We inserted the HJ's sinusoidal phase curve into a 1470 day, 30 minute cadence light curve with 200\footnote{200 ppm is the median long-cadence photometric precision as calculated using KOI host stars with $12.5 <$ KepMag $<13.5$.} ppm precision. This corresponds to 17 quarters of \textit{Kepler} long cadence photometry on a typical V $\sim$ 13 mag star. In a Lomb-Scargle (LS) periodogram \citep{1976ApSS..39..447L, 1982ApJ...263..835S} of the light curve, over a 1-6 day range, the 3.5 day signal is easily the highest peak. A Gaussian fit to the highest peak in the periodogram yields $P_J = 3.5001 \pm 0.003$. A least-squares sine fit to the phase-folded light curve, binned such that 400 points span the orbit, returns an amplitude of $\sim$36.3 ppm, close to the 36 ppm input. The recovered phase curve is shown in Figure ~\ref{TestCase10_PhaseTTV}.


\begin{figure}
\epsscale{1.2}
\plotone{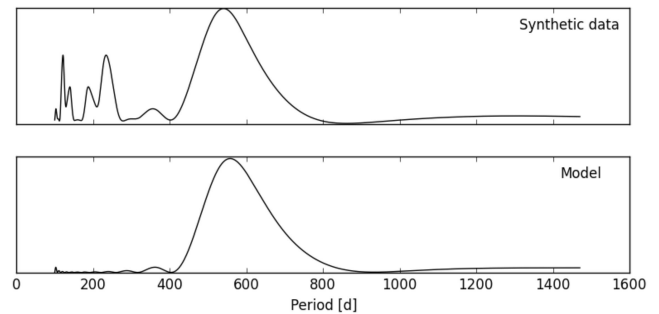}
\caption{The LS periodograms of the TTVs for the synthetic data (top) and model TTVs (bottom). Each features a peak close to 560 days.}
\label{TestCase10_LS}
\end{figure}

The astrometric TTVs of the 3 $R_{\rm{\earth}}$ planet form a sinusoidal oscillation with amplitude 2.17 min and period 560 days. This is not easy to detect given the expected 4.2 min\footnote{4.2 min is the median of the TTV median uncertainties for Kepler KOIs with $12.5 <$ KepMag $<13.5$ and $0.0008 < (R_P/R_{\star})^2 < 0.0012$.} median uncertainties that are typical for \textit{Kepler} TTVs of a 3 $R_{\rm{\earth}}$ planet orbiting a V $\sim$ 13 mag, 1 $R_{\rm{\sun}}$ star. Figure ~\ref{TestCase10_PhaseTTV} shows the expected TTVs with scatter consistent with typical observational uncertainty.  When comparing the TTV model with data, the TTV phase may be strongly constrained using the epoch of the phase curve maximum. Although a visual comparison between the TTV model and simulated data is suggestive at best, the LS periodograms of the model and data show a strong correspondence (Figure \ref{TestCase10_LS}). Each periodogram has a peak near 560 days. 

The phase curve and astrometric TTVs of this fiducial system are both readily detectable given \textit{Kepler}-quality photometry.  Although neither signal would itself yield a conclusive detection, the combination of the two, coupled with the demand of phase correlation, is extremely powerful. Promising candidates, furthermore, can readily be confirmed using RV observations.

\section{CANDIDATE NON-TRANSITING HOT JUPITERS IN THE \textit{Kepler} DATA}

We performed a brief, non-comprehensive assessment of the archived \textit{Kepler} data to search for non-transiting HJs using the ``phase+astrometric TTV'' technique. We examined a subset of targets flagged as confirmed exoplanets or KOIs from the MAST \textit{Kepler} data archive\footnote{\url{https://archive.stsci.edu/Kepler/}}. We used the publicly available light curve files containing the pre-search data conditioning (PDC) simple aperture photometry \citep{2012PASP..124..985S}. We used \textit{Kepler} Q1-Q17 long- and short- cadence photometric data and Q1-Q17 TTV data from \cite{Holczer2016}. 

We first searched the light curves for phase curve detections. We filtered the light curves by removing variability on timescales greater than 6 days using the \textit{kepflatten} routine in the \textit{PyKE Kepler} data reduction software \citep{2012ascl.soft08004S}. We then stitched the light curves from various quarters and cadence modes together. 

Operating on these detrended and concatenated light curves, we removed 3$\sigma$ outliers and calculated each target's Lomb-Scargle (LS) periodogram in a 1 to 6 days period range. We folded the light curve according to the peak period in the periodogram and performed a least-squares sinusoidal fit to the resulting phase curve. The phase of the fit is used to derive a time epoch at which the HJ is directly behind the star, to within uncertainties caused by possible shifts between the brightest region on the planet and the sub-stellar point \citep[as discussed in][]{2015AJ....150..112S}. At this stage, many of the target light curves show roughly sinusoidal phase curve variations. Some are likely false positives due to contamination from stellar variability. The TTVs help rule out these false positives. An autocorrelation technique could be another effective method. 

For each \textit{Kepler} confirmed planet/KOI in our selection, we generated a profile of the expected TTVs to compare them to the \cite{Holczer2016} observed TTVs. In the TTV calculations, we used the period of the HJ as detected from the phase curve, the average period of the planet/KOI as detected from its transits and reported at the NASA Exoplanet Archive\footnote{\label{ExoArch}\url{http://exoplanetarchive.ipac.caltech.edu}}, the stellar mass estimated from stellar parameters and the planetary transit model fitting, a fiducial HJ mass of 1 ${M_{\rm {Jup}}}$, and the time epochs of the KOI transit and the phase curve maximum.

We visually compared each candidate's modeled and observed TTVs and simultaneously examined the candidates' detected phase curves. We also compared the LS periodograms of the expected and observed TTVs. We established a list of candidates showing both detectable phase curve variations and TTVs for which the data and model qualitatively matched within observational uncertainty. 

\section{A candidate non-transiting hot Jupiter in the KOI-1822 system}

Here we present one of our detected candidates, a possible non-transiting HJ orbiting \textit{Kepler} star 5124667. This star has $T_{\rm{eff}}=5504 \pm 87 $ K, metallicity Fe/H $= 0.30 \pm 0.15$, mass 1.099$\substack{+0.098 \\ -0.131}$  $M_{\rm{\sun}}$ and radius 1.737$\substack{+0.26 \\ -0.608}$ $R_{\rm{\sun}}$ \citep[][Q1-Q17 DR25]{2014ApJS..211....2H}. It hosts KOI-1822.01, a 3.16$\substack{+1.19 \\ -0.65}$ $R_{\rm{\earth}}$ candidate planet in a $150.87799\pm 0.000384$ day orbit, as reported on the NASA Exoplanet Archive\textsuperscript{\ref{ExoArch}} \citep{2013PASP..125..989A}.

\begin{figure}
\epsscale{1.2}
\plotone{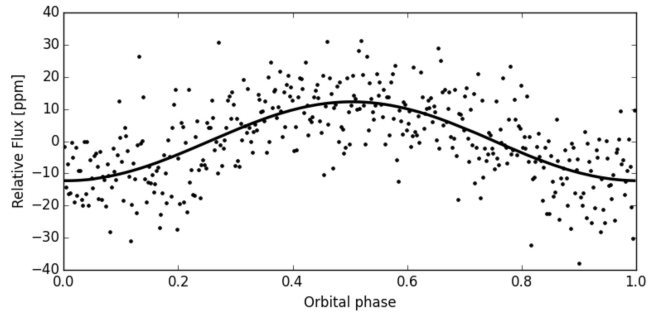}
\caption{The detected phase curve for the candidate non-transiting HJ in the KOI-1822 system. The light curve was phase-folded at the post-MCMC period estimate, 3.908 days, and median binned such that 400 points span the orbit. The black line is a fixed-period, least-squares sinusoidal fit.}
\label{KOI1822_PhaseCurve}
\end{figure}

\begin{figure}
\epsscale{1.2}
\plotone{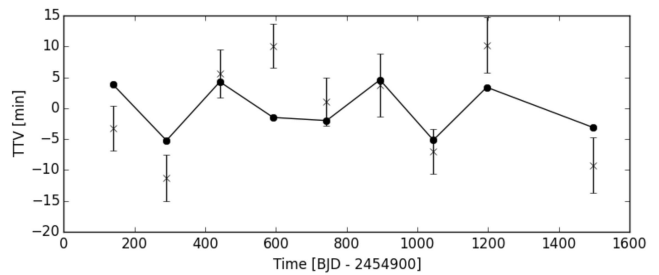}
\caption{The MCMC derived fit to KOI-1822.01's observed TTVs. The ``x" symbols represent the data, and the black line is the fit using the parameters of the posterior means.}
\label{KOI1822_Fit}
\end{figure}

Using all available light curve data, the LS periodogram returns a 13 ppm amplitude phase curve detection with period $3.9105 \pm 0.005$ days. The phase curve and a sinusoidal fit are shown in Figure \ref{KOI1822_PhaseCurve}. Moreover, \cite{Holczer2016} measured significant TTVs for KOI-1822.01 (Figure \ref{KOI1822_Fit}). To evaluate the consistency between KOI-1822.01's observed TTVs and those expected from a non-transiting HJ, we performed parameter estimation using the Markov Chain Monte Carlo (MCMC) Metropolis-Hastings algorithm. The fit did not involve the phase curve data, but rather utilized the detected period as a prior for $P_J$. Given the large uncertainties and paucity of data points in the TTVs, we caution the reader to interpret the following parameter estimation as little more than a plausibility argument.

For simplicity we assumed circular orbits, so the MCMC contained 8 free parameters: $M_{\rm{\star}}$, $M_J$, $P_J$, $P_p$, $I_J$, $\Omega_J$, the transit epoch $t_{\rm{0}}$, and the epoch of the HJ's superior conjunction $t_J$. Most of these parameters have very tight priors. We used Gaussian priors with means and standard deviations for $M_{\rm{\star}}$, $P_p$, and $t_{\rm{0}}$ derived from the transit model fit as reported on the NASA Exoplanet Archive\textsuperscript{\ref{ExoArch}}. The prior distribution for $P_J$ was derived from a Gaussian fit to the peak period in the LS periodogram. The prior mean for $t_J$ was estimated from the sinusoidal fit to the phase curve, and the standard deviation was taken to be 0.3 days. The priors for $M_J$ and $\Omega_J$ were 2.0 $\pm$ 0.5 $M_{\rm{Jup}}$ and 0 $\pm$ 120$^\circ$, respectively, and $I_J$ was a half-Gaussian with standard deviation 30$^\circ$.

The likelihood of the TTV data $\bf{d}$ given the set $\bm{\theta}$ of model parameters is given by $L\left(\bf{d} | \bm{\theta}\right) \propto e^{-\chi^2/2}$ where $\chi^2$ is the chi-square error. We used the Metropolis-Hastings algorithm to sample from the posterior distribution given these priors and likelihoodays. The TTV model fit using the best-fit parameters (posterior means) after 1000000 samples is presented in Figure \ref{KOI1822_Fit}. 

The posterior means of  $M_{\rm{\star}}$, $P_p$, and $t_{\rm{0}}$ are consistent with the prior means, as they should be. The best-fit estimates for the HJ's parameters are as follows: $P_J = 3.908 \pm 0.003$ days, $I_J = 83 \substack{+7 \\ -17} ^\circ$, $\Omega_J = 16 \pm 99^\circ$, and $M_J = 2.0\pm 0.44$ $M_{\rm{Jup}}$. Using the mass/radius distribution of known HJs with P $<$ 5 days, the HJ mass estimate is consistent with a radius of $1.3 \pm 0.3$ $R_{\rm{Jup}}$. For consistency with the 13 ppm phase curve amplitude, the HJ's geometric albedo should be $\sim$0.09, which is a physically reasonable estimate. Moreover, the best-fit estimate for $t_J$ is consistent with the epoch of the phase curve maximum, meaning the two independent measurements are phase correlated. 

Although TDVs were detected for KOI-1822.01 by \cite{Holczer2016}, the observational uncertainties were too large for a rigorous examination here. It should be noted, however, that the observed TDVs are not in disagreement with the expected signal due to a non-transiting HJ. 

Finally, the large stellar metallicity, Fe/H $= 0.30 \pm 0.15$ \citep[][Q1-Q17 DR25]{2014ApJS..211....2H}, further increases the likelihood of the candidate HJ's existence, given the well-known correlation between stellar metallicity and giant planet occurrence \citep{2004AA...415.1153S, 2005ApJ...622.1102F}.

Though the preceding analysis constitutes a plausibility argument, this candidate non-transiting HJ is readily RV-confirmable given its magnitude (KepMag = 12.4) and its large expected Doppler half-amplitude (K $\sim$ 250 m/s). This is easily detectable on a telescope like the Automated Planet Finder, which would attain $\sim$ 8 m/s precision in a 1 hr measurement. Moreover, the estimates of $P_J$ and the epoch of superior conjunction, $t_J$, can be used to estimate the quadrature ephemeris. If the HJ is present, the combined phase curve, TTV, TDV, and RV data may enable a full description of the candidate HJ's orbit.

\section{CONCLUSION}

We have shown that the combination of full phase light curves and astrometric transit timing variations generates an effective method for identifying candidate non-transiting HJs in multiple-planet systems, and as a proof of concept, we identified a candidate in the Kepler 5124667/KOI-1822 system. If we assume that $\sim$3,000 stars among the $\sim$150,000 monitored by \textit{Kepler} will be confirmed to harbor transiting super-Earths, and if we assume that in situ formation is a significant channel for creating HJs (which have an intrinsic occurrence fraction of $\sim$0.5\%), then we expect that $\sim$10 non-transiting HJs can be identified using the method outlined here and confirmed using quick-look Doppler spectroscopy. It also bears mentioning that photometric data from the K2 and TESS Missions will be equally well suited to identifying such systems.
\\

We acknowledge support from the NASA Astrobiology Institute through a cooperative agreement between NASA Ames Research Center and the University of California at Santa Cruz, and from the NASA TESS Mission through a cooperative agreement between M.I.T. and UCSC. We thank Darin Ragozzine for providing access to  Q1-Q17 KOI TTV data. 
 

\begin{thebibliography}{}


\bibitem[{{Agol} {et~al.}(2005){Agol}, {Steffen}, {Sari}, \&
  {Clarkson}}]{2005MNRAS.359..567A}
{Agol}, E., {Steffen}, J., {Sari}, R., \& {Clarkson}, W. 2005, \mnras, 359, 567

\bibitem[{{Akeson} {et~al.}(2013){Akeson}, {Chen}, {Ciardi}, {Crane}, {Good},
  {Harbut}, {Jackson}, {Kane}, {Laity}, {Leifer}, {Lynn}, {McElroy}, {Papin},
  {Plavchan}, {Ram{\'{\i}}rez}, {Rey}, {von Braun}, {Wittman}, {Abajian},
  {Ali}, {Beichman}, {Beekley}, {Berriman}, {Berukoff}, {Bryden}, {Chan},
  {Groom}, {Lau}, {Payne}, {Regelson}, {Saucedo}, {Schmitz}, {Stauffer},
  {Wyatt}, \& {Zhang}}]{2013PASP..125..989A}
{Akeson}, R.~L., {Chen}, X., {Ciardi}, days., {et~al.} 2013, \pasp, 125, 989

\bibitem[{{Angerhausen} {et~al.}(2015){Angerhausen}, {DeLarme}, \&
  {Morse}}]{2015PASP..127.1113A}
{Angerhausen}, days., {DeLarme}, E., \& {Morse}, J.~A. 2015, \pasp, 127, 1113

\bibitem[{{Armstrong} {et~al.}(2013){Armstrong}, {Martin}, {Brown}, {Faedi},
  {G{\'o}mez Maqueo Chew}, {Mardling}, {Pollacco}, {Triaud}, \&
  {Udry}}]{2013MNRAS.434.3047A}
{Armstrong}, days., {Martin}, days.~V., {Brown}, G., {et~al.} 2013, \mnras, 434, 3047

\bibitem[{{Batygin} {et~al.}(2015){Batygin}, {Bodenheimer}, \&
  {Laughlin}}]{2015arXiv151109157B}
{Batygin}, K., {Bodenheimer}, P.~H., \& {Laughlin}, G.~P. 2015, ArXiv e-prints,
  arXiv:1511.09157

\bibitem[{{Beaug{\'e}} \& {Nesvorn{\'y}}(2012)}]{2012ApJ...751..119B}
{Beaug{\'e}}, C., \& {Nesvorn{\'y}}, days. 2012, \apj, 751, 119

\bibitem[{{Becker} {et~al.}(2015){Becker}, {Vanderburg}, {Adams}, {Rappaport},
  \& {Schwengeler}}]{2015ApJ...812L..18B}
{Becker}, J.~C., {Vanderburg}, A., {Adams}, F.~C., {Rappaport}, S.~A., \&
  {Schwengeler}, H.~M. 2015, \apjl, 812, L18

\bibitem[{{Boley} {et~al.}(2016){Boley}, {Granados Contreras}, \&
  {Gladman}}]{2016ApJ...817L..17B}
{Boley}, A.~C., {Granados Contreras}, A.~P., \& {Gladman}, B. 2016, \apjl, 817,
  L17

\bibitem[{{Coughlin} \& {L{\'o}pez-Morales}(2012)}]{2012AJ....143...39C}
{Coughlin}, J.~L., \& {L{\'o}pez-Morales}, M. 2012, \aj, 143, 39

\bibitem[{{Esteves} {et~al.}(2013){Esteves}, {De Mooij}, \&
  {Jayawardhana}}]{2013ApJ...772...51E}
{Esteves}, L.~J., {De Mooij}, E.~J.~W., \& {Jayawardhana}, R. 2013, \apj, 772,
  51

\bibitem[{{Esteves} {et~al.}(2015){Esteves}, {De Mooij}, \&
  {Jayawardhana}}]{2015ApJ...804..150E}
---. 2015, \apj, 804, 150

\bibitem[{{Faigler} \& {Mazeh}(2011)}]{2011MNRAS.415.3921F}
{Faigler}, S., \& {Mazeh}, T. 2011, \mnras, 415, 3921

\bibitem[{{Faigler} {et~al.}(2012){Faigler}, {Mazeh}, {Quinn}, {Latham}, \&
  {Tal-Or}}]{2012ApJ...746..185F}
{Faigler}, S., {Mazeh}, T., {Quinn}, S.~N., {Latham}, days.~W., \& {Tal-Or}, L.
  2012, \apj, 746, 185

\bibitem[{{Fischer} \& {Valenti}(2005)}]{2005ApJ...622.1102F}
{Fischer}, days.~A., \& {Valenti}, J. 2005, \apj, 622, 1102

\bibitem[{{Gibson} {et~al.}(2009){Gibson}, {Pollacco}, {Simpson}, {Barros},
  {Joshi}, {Todd}, {Keenan}, {Skillen}, {Benn}, {Christian}, {Hrudkov{\'a}}, \&
  {Steele}}]{2009ApJ...700.1078G}
{Gibson}, N.~P., {Pollacco}, days., {Simpson}, E.~K., {et~al.} 2009, \apj, 700,
  1078
  
\bibitem[Holczer {et al.}(2016)]{Holczer2016}
Holczer, T., Mazeh, T., Nachmani, G., et al. 2016, ApJ, in press

\bibitem[{{Holman} \& {Murray}(2005)}]{2005Sci...307.1288H}
{Holman}, M.~J., \& {Murray}, N.~W. 2005, Science, 307, 1288

\bibitem[{{Huang} {et~al.}(2016){Huang}, {Wu}, \&
  {Triaud}}]{2016arXiv160105095H}
{Huang}, C.~X., {Wu}, Y., \& {Triaud}, A.~H.~M.~J. 2016, ArXiv e-prints,
  arXiv:1601.05095

\bibitem[{{Huber} {et~al.}(2014){Huber}, {Silva Aguirre}, {Matthews},
  {Pinsonneault}, {Gaidos}, {Garc{\'{\i}}a}, {Hekker}, {Mathur}, {Mosser},
  {Torres}, {Bastien}, {Basu}, {Bedding}, {Chaplin}, {Demory}, {Fleming},
  {Guo}, {Mann}, {Rowe}, {Serenelli}, {Smith}, \&
  {Stello}}]{2014ApJS..211....2H}
{Huber}, days., {Silva Aguirre}, V., {Matthews}, J.~M., {et~al.} 2014, \apjs, 211,
  2

\bibitem[{{Kley} \& {Nelson}(2012)}]{2012ARAA..50..211K}
{Kley}, W., \& {Nelson}, R.~P. 2012, \araa, 50, 211

\bibitem[{{Latham} {et~al.}(2011){Latham}, {Rowe}, {Quinn}, {Batalha},
  {Borucki}, {Brown}, {Bryson}, {Buchhave}, {Caldwell}, {Carter},
  {Christiansen}, {Ciardi}, {Cochran}, {Dunham}, {Fabrycky}, {Ford}, {Gautier},
  {Gilliland}, {Holman}, {Howell}, {Ibrahim}, {Isaacson}, {Jenkins}, {Koch},
  {Lissauer}, {Marcy}, {Quintana}, {Ragozzine}, {Sasselov}, {Shporer},
  {Steffen}, {Welsh}, \& {Wohler}}]{2011ApJ...732L..24L}
{Latham}, days.~W., {Rowe}, J.~F., {Quinn}, S.~N., {et~al.} 2011, \apjl, 732, L24

\bibitem[{{Lomb}(1976)}]{1976ApSS..39..447L}
{Lomb}, N.~R. 1976, \apss, 39, 447

\bibitem[{{Placek} {et~al.}(2014){Placek}, {Knuth}, \&
  {Angerhausen}}]{2014ApJ...795..112P}
{Placek}, B., {Knuth}, K.~H., \& {Angerhausen}, days. 2014, \apj, 795, 112

\bibitem[{{Santos} {et~al.}(2004){Santos}, {Israelian}, \&
  {Mayor}}]{2004AA...415.1153S}
{Santos}, N.~C., {Israelian}, G., \& {Mayor}, M. 2004, \aap, 415, 1153

\bibitem[{{Scargle}(1982)}]{1982ApJ...263..835S}
{Scargle}, J.~days. 1982, \apj, 263, 835

\bibitem[{{Shporer} \& {Hu}(2015)}]{2015AJ....150..112S}
{Shporer}, A., \& {Hu}, R. 2015, \aj, 150, 112

\bibitem[{{Shporer} {et~al.}(2011){Shporer}, {Jenkins}, {Rowe}, {Sanderfer},
  {Seader}, {Smith}, {Still}, {Thompson}, {Twicken}, \&
  {Welsh}}]{2011AJ....142..195S}
{Shporer}, A., {Jenkins}, J.~M., {Rowe}, J.~F., {et~al.} 2011, \aj, 142, 195

\bibitem[{{Steffen} {et~al.}(2012){Steffen}, {Ragozzine}, {Fabrycky}, {Carter},
  {Ford}, {Holman}, {Rowe}, {Welsh}, {Borucki}, {Boss}, {Ciardi}, \&
  {Quinn}}]{2012PNAS..109.7982S}
{Steffen}, J.~H., {Ragozzine}, days., {Fabrycky}, days.~C., {et~al.} 2012,
  Proceedings of the National Academy of Science, 109, 7982

\bibitem[{{Still} \& {Barclay}(2012)}]{2012ascl.soft08004S}
{Still}, M., \& {Barclay}, T. 2012, {PyKE: Reduction and analysis of Kepler
  Simple Aperture Photometry data}, Astrophysics Source Code Library,
  ascl:1208.004

\bibitem[{{Stumpe} {et~al.}(2012){Stumpe}, {Smith}, {Van Cleve}, {Twicken},
  {Barclay}, {Fanelli}, {Girouard}, {Jenkins}, {Kolodziejczak}, {McCauliff}, \&
  {Morris}}]{2012PASP..124..985S}
{Stumpe}, M.~C., {Smith}, J.~C., {Van Cleve}, J.~E., {et~al.} 2012, \pasp, 124,
  985

\bibitem[{{Wu} \& {Murray}(2003)}]{2003ApJ...589..605W}
{Wu}, Y., \& {Murray}, N. 2003, \apj, 589, 605

\end{thebibliography}

\end{document}